\documentstyle[amssymb,12pt]{article}

\begin{document}

\baselineskip=0.7cm
\begin{center}
{\Large\bf A BIGRADED VERSION OF\\
 THE WEIL ALGEBRA AND OF\\
 THE WEIL HOMOMORPHISM\\
 FOR DONALDSON INVARIANTS}
\end{center}
\vspace{0.75cm}
\begin{center}
{\large Elementary algebra and cohomology\\
behind the Baulieu-Singer approach\\
to Witten's topological Yang-Mills\\
quantum field theory}\\
\end{center}
\vspace{0.75cm}
\begin{center}
{\large Michel DUBOIS-VIOLETTE}
\end{center}
\vspace{0.2cm}
\begin{center}
{\small\it Laboratoire de Physique Th\'eorique\footnote{Laboratoire associ\'e
au
C.N.R.S.}\\ B\^atiment 211, Universit\'e Paris XI\\
91405 ORSAY Cedex}\\

E-mail: flad@qcd.th.u-psud.fr\\

\today
\end{center}

\vspace {1cm}
\begin{abstract}
We describe a bigraded generalization of the Weil algebra, of its basis
and of the characteristic homomorphism which besides ordinary
characteristic classes also maps  on Donaldson invariants.
\end{abstract}
\vspace{1cm}

L.P.T.H.E.-ORSAY 94/09
\newpage
\section*{Introduction}

In [1], a small differential algebra was introduced in connection with
gauge fixing \`a la B.R.S., [2], in Witten's topological quantum field
theory [11]. Furthermore, in the last part of [1],$\S$5, this algebra
was identified with a differential subalgebra of the algebra of
differential forms on the product $P\times {\cal C}$ of a principal
bundle $P$ over a manifold $M$ with the space $\cal C$ of principal
connections of $P$. By a slight ``abstraction" one can produce an
algebra $\frak A$ which is a free bigraded-commutative differential
algebra which admits the above quoted differential algebra (of [1]) as
homomorphic image. This algebra, which is described in part I of the
present paper, is obviously a bigraded version of the Weil algebra, [10]
(see also in [4] and [7]) ; it is a contractible algebra. Therefore the
various cohomologies of $\frak A$ as well as the ones of its images are
trivial. It was claimed in [9] that the relevant cohomologies where
the basic cohomologies for operations in the sense of H. Cartan [4],[7].
Actually this is also implicit in [1] since there, it is claimed, (in
$\S$5), that the relevant cohomology is the de Rham cohomology of
$M\times {\cal C}/{\cal G}$ (and not the one of $P\times {\cal C})$ where
$\cal G$ is the group of gauge transformations of $P$. It is the aim of
this paper to show that, by using $\frak A$ as generalization of the Weil
algebra, one can generalize the Weil homomorphism in such a way that
beside the usual characteristic classes, it makes contact with the
Donaldson invariants [5]. That there is connection between the usual
characteristic classes and the Donaldson invariant is natural in view of
the apparences of the Donaldson-Witten cocycles [11] which are bigraded
expansions of the Chern-Weil cocycles. It is worth noticing here that
$\frak A$ is a generalization of the Weil algebra based on the finite
dimensional Lie algebra of the structure group of $P$ and {it not} the
one based on the infinite dimensional Lie algebra of the group of gauge
transformations of $P$, [8]; in particular, $\frak A$ is finitely
generated.\\
The plan of the paper is the following. In part I we describe the
bigraded generalization of the Weil algebra and its basis and we compute
its various basic cohomologies. In part II we describe the corresponding
appropriate generalization of the characteristic homomorphism. One can
remark that the independence of the metric in [1] follows from the more
general fact proved here that the characteristic homomorphism does not
depend on the chosen connection of the space of connection over $P$ (the
structure group being there the gauge group). For a review on
topological quantum field theory as well as for detailed references on
the subject we refer to the Physics Report [3]. We apologize on the fact
that we denote by $\alpha$ or more precisely by $\underline{\alpha}$
what is almost everywhere in the literature denoted by $c$,  e.g. [1,3].

\newpage
 \begin{center}
\large Part I\\
{\bf A UNIVERSAL MODEL}\\
Generalisation of the Weil algebra
\end{center}

\vspace{1cm}

\section{Description of the model}

Let $\frak g$ be a finite dimensional Lie algebra and let
($E_i$) be a basis of $\frak g$. Consider eight copies
$\frak g^\ast_A, \frak g^\ast_F, \frak g^\ast_\alpha,
\frak g^\ast_\varphi, \frak g^\ast_\psi, \frak
g^\ast_\xi, \frak g^\ast_B$ and $\frak g^\ast_\beta$ of
the dual $\frak g^\ast$ of $\frak g$ with dual basis
respectively denoted by $(A^i), (F^i), (\alpha^i),
(\varphi^i), (\psi^i), (\xi^i), (B^i)$ and $(\beta^i)$.
Let $\frak A(\frak g)$, (or simply $\frak A$), be the
free graded commutative algebra generated by the $A^i$
and $\alpha^i$ in degree one, the $F^i$, $\varphi^i$,
$\psi^i$ and $\xi^i$ in degree two and the $B^i$ and
$\beta^i$ in degree three. On the space $\frak g \otimes
\frak A$, there is a natural bracket $[\bullet, \bullet]$
defined by $[X\otimes P, Y\otimes Q] = [X,Y]\otimes P\cdot
Q$, for $X,Y\in \frak g$ and $P,Q\in \frak A$. For any
linear mapping $R$ of $\frak A$ in itself one defines a
linear mapping, again denoted by $R$, of $\frak g \otimes
\frak A$ in itself by $R(X\otimes P)=X\otimes R(P)$ for
$X\in \frak g$ and $P\in \frak A$. Let us introduce
the following elements of $\frak g \otimes \frak A$
associated to the generators of $\frak A$:
$A=E_i\otimes A^i$, $F = E_i\otimes F^i$, $\alpha = E_i
\otimes \alpha^i$, $\varphi = E_i\otimes \varphi^i$, $\psi =
E_i \otimes \psi^i$, $\xi = E_i\otimes \xi^i$, $B=E_i\otimes
B^i$ and $\beta = E_i\otimes \beta^i$. With the above
notations, one has the following lemma.

\newtheorem{lemma}{LEMMA}[section]

\begin{lemma}
There are unique antiderivations $d$ and $\delta$ of
$\frak A$ satisfying\linebreak[4] $d^2=0$, $\delta^2=0$ and
$d\delta + \delta d = 0$ such that $F=dA +
\frac{1}{2}[A,A]$, $\varphi=\delta \alpha +
\frac{1}{2}[\alpha,\alpha]$, $\psi=\delta A + d\alpha +
[A,\alpha]$, $\xi = \delta A$, $B=\delta F + [\alpha ,F]$
and $\beta = d\varphi + [A,\varphi]$.
\end{lemma}

\noindent {\bf Proof}. By a change of generators,
$\frak A$ is the free graded commutative algebra
generated by the $A^i$ and $\alpha^i$ in degree one,
$dA^i$, $\delta A^i$, $d\alpha^i$, $\delta \alpha^i$ in
degree two and $\delta d A^i$, $d\delta \alpha^i$ in
degree three. The lemma is thus obvious and $\frak A$ is
contractible for $d$ and for $\delta$.$\square$\\
One defines an underlying bigraduation on $\frak A$ by
giving to the $A^i$ the bidegree (1,0), to the $F^i$ the
bidegree (2,0), to the $\alpha^i$ the bidegree (0,1), to
the $\varphi^i$ the bidegree (0,2), to the $\psi^i$ and
the $\xi^i$ the bidegree (1,1), to the $B^i$ the bidegree
(2,1) and to the $\beta^i$ the bidegree (1,2). $\frak A$
is then a bigraded differential algebra with total
differential $d+\delta$, $d$ being the part of bidegree
(1,0) and $\delta$ the part of bidegree (0,1). Since
$\frak A$ is contractible for each of these differentials
its cohomologies are trivial $(H^{k,\ell}(\frak A,d)=0$ if
$k+\ell>0$, $H^{0,0}(\frak A,d)\cong \Bbb C$ , etc.). The
interest of this bigraded differential algebra lies in
the following universal property which is closely related
to the triviality of its cohomology.

\begin{lemma}
 Let $\Omega$ be a
bigraded commutative differential algebra with
differential $d+\delta$, $d$ of bidegree (1,0) and
$\delta$ of bidegree (0,1). Let $\underline{A}=E_i\otimes
\underline{A}^i$ be an element of $\frak g \otimes
\Omega^{1,0}$
 and $\underline{\alpha}=E_i\otimes \underline{\alpha}^i$ be an element of
$\frak g\otimes \Omega^{0,1}$. Then there is a unique
homomorphim of bigraded differential algebras $h:\frak A
\rightarrow \Omega$ such that $h(A^i)=\underline{A}^i$ and
$h(\alpha^i)=\underline{\alpha}^i$, (in short
$h(A)=\underline{A}, h(\alpha)=\underline{\alpha}$).
\end{lemma} Notice that $\frak A$ contains two Weil algebras
namely $W_d(\frak g)=(\Lambda\frak g^\ast_A\otimes S\frak
g^\ast_F,d)$ and $W_\delta(\frak g)=(\Lambda \frak
g^\ast_\alpha \otimes S\frak g^\ast_\varphi,\delta)$. The
first one, $W_d(\frak g)$, will play in the sequel the role of
the Weil algebra (see in part II); its base (i.e. basic
subalgebra) consists of the $\cal I(F)$ where $\cal I$ runs
over all $ad^\ast(\frak g)$-invariant elements of $S\frak
g^\ast$ (i.e. invariant polynomials on $\frak g$). Let $\frak
B(\frak g)$, (or simply $\frak B$), be the set of elements of
$\frak A$ of the form $\cal I(F,\psi,\varphi, B,\beta)$ where
$\cal I$ runs over all $ad^\ast(\frak g)$-invariant elements
of $S\frak g^\ast \otimes S\frak g^\ast \otimes S\frak g^\ast
\otimes \Lambda \frak g^\ast \otimes \Lambda\frak g^\ast$,
(invariant ``polynomials" in $F,\psi,\varphi, B$ and
$\beta$). $\frak B$ is a bigraded subalgebra of $\frak A$
which is easily shown to be stable by $d$ and by
$\delta$. For reasons which will become clear in part II,
the bigraded differential algebra $\frak B$ will be
called {\it the base of} $\frak A$.\\
$\frak B$ is in fact the set of elements which are basic for two
operations of $\frak g$ in $\frak A$ which we now describe. One defines
first an operation of $\frak g$ in $\frak A$, $X\mapsto i_X$, which
extends the usual operation of $\frak g$ in the Weil algebra $W_d(\frak
g)$ by setting, for $X\in \frak g$, $i_X(A)=X$ and $i_X$(other
generators)=0 which implies $i_X\delta + \delta i_X=0$ and, with
$L_X=i_Xd+di_X$, $L_X(Z)=[Z,X]$ for all generators
$Z=A,F,\alpha,\varphi,\psi,\xi, B,\beta$. By interchanging the bidegrees,
one defines similarily an other operation of $\frak g$ in $\frak A$,
$X\mapsto i'_X$, which extends the one of $W_\delta(\frak g)$. Namely one
sets, for $X\in \frak g$, $i'_X(\alpha)=X$,
$i'_X(A)=i'_X(F)=i'_X(\varphi)=i'_X(\psi)=i'_X(B)=i'_X(\beta)=0$ {\it
and} $i'_X(d\alpha)=0$ (notice that then $i'_X(\delta
A)=i'_X(\xi)\neq 0)$; this implies $i'_Xd +di'_X=0$ and, with
$L'_X=i'_X\delta + \delta i'_X$, $L'_X(Z)=[Z,X]$ for all
generators $Z=A,F,\alpha,\varphi,\psi,\xi,B, \beta$. One verifies that
one has:
$$\frak B = \cap \{\ker (i_X)\cap \ker (L_Y) \cap \ker
(i'_{X'}) \cap \ker (L'_{Y'})\} X,Y,X',Y' \in \frak g.$$
However, in part II, only the first operation will be mapped
homorphically on a similar operation of $\frak g$. Nevertheless $\frak
B$ will be mapped into a basis of two operations.\\
In constrast to the case of $\frak A$, the various
cohomologies of $\frak B$ are non-trivial and it is the
aim of the next sections of part I to compute these
cohomologies.

\section {The $d$ and the $\delta$ cohomologies of
$\frak B(\frak g)$.}

Let $d_1,d_2,\delta_1$ and $\delta_2$ be the unique
antiderivations of $\frak A$ satisfying:
$$d_1\psi=-B, d_1\varphi=\beta, d_1(\mbox{other
generators})=0;$$
$$d_2B =-[F,\psi], d_2\beta = [F,\varphi],
d_2(\mbox{other generators})=0;$$
$$\delta_1F=B, \delta_1\psi=-\beta,\delta_1 (\mbox{other
generators})=0;$$
 $$\delta_2B=[\varphi,F],
\delta_2\beta=-[\varphi,\psi], \delta_2 (\mbox{other
generators})=0.$$
 $\frak B$ is stable by $d_1,d_2,\delta_1$ and
$\delta_2$ and one has the following lemma.

\begin{lemma}
The restriction of $d$ to $\frak B$ coincides with the
restriction of $d_1+d_2$ and the restriction of $\delta$
to $\frak B$ coincides with the restriction of
$\delta_1+\delta_2$.

\end{lemma}
\noindent {\bf Proof}. One has
$(d+\delta)(A+\alpha)+\frac{1}{2}[A+\alpha,A+\alpha]=F+\psi+\varphi$
which implies
$(d+\delta)(F+\psi+\varphi)+[A+\alpha,F+\psi+\varphi]=0$.
By taking the homogeneous components and by using the
definitions, one obtains:

\[ \left\{ \begin{array}{ll}
dF +  [A,F] &= 0\\
d\psi  +  [A,\psi] & = -B \\
d\varphi + [A,\varphi] &= \beta \\
dB + [A,B] & = - [F,\psi] \\
d\beta + [A,\beta] & = [F,\varphi]
\end{array}
\right.
\mbox{and}\> \> \left\{ \begin{array}{ll}
\delta F + [\alpha,F] &= B \\
\delta \psi + [\alpha,\psi] &= -\beta \\
\delta \varphi + [\alpha,\varphi] &= 0 \\
\delta B + [\alpha,B] &= [\varphi, F] \\
\delta \beta + [\alpha,\beta] &= -[\varphi, \psi].
\end{array}
\right.
\]
Thus the lemma would be obvious without the terms
$[A,\cdot]$ and the terms $[\alpha,\cdot]$. However if
$\cal I(F,\psi,\varphi,B,\beta)\in \frak B$, it follows
from the invariance of $\cal I$ that one can enter $d$ in
$\cal I$ by replacing it everywhere by $d+[A,\cdot]$ and
that one can enter $\delta$ in $\cal I$ by replacing it
everywhere by $\delta+[\alpha,\cdot]$. This implies the
result.~$\square$

\begin{lemma}
Let $d'_1$ and $\delta'_1$ be the unique antiderivations
of $\frak A$ satisfying:
$$d'_1B=-\psi,
d'_1\beta=\varphi, d'_1(\mbox{other generators})=0$$ and
$$\delta'_1B=F, \delta'_1\beta=-\psi, \delta'_1(\mbox{other
generators})=0.$$ Then one has on $\frak B$: $dd'_1+d'_1d$ =
degree in $(\psi,\varphi,B,\beta)$ and
$\delta\delta'_1+\delta'_1\delta$=degree in
$(F,\psi,B,\beta)$.
\end{lemma}

\noindent{\bf Proof}. One has on $\frak A$:\\
$$d_1d'_1+d'_1d_1 = \mbox{degree in}
(\psi,\varphi,B,\beta), \> d_2 d'_1+d'_1d_2 = 0$$
$$\delta_1\delta'_1 + \delta'_1\delta_1 = \mbox{degree
in} (F,\psi,B,\beta), \> \delta_2\delta'_1 +
\delta'_1\delta_2 = 0.$$\\
Since both sides of these equations are derivations of
$\frak A$, it is sufficient to verify that they are true
on the generators which is straightforward. The lemma
follows then from lemma 2.1. $\square$\\
We are now ready to describe the $d$ and $\delta$
cohomologies of $\frak B$. Let $\cal I_S(\frak g)$ be the
space of invariant polynomials on $\frak g$ and let $\cal
I^n_S(\frak g)$ be the subspace of homogeneous invariant
polynomials of degree $n$ on $\frak g$; one has the following
result.

\newtheorem{theorem}{THEOREM}[section]

\begin{theorem}
The $d$ and the $\delta$ cohomologies of $\frak B$ are
given by:
\end{theorem}

\[ \left\{ \begin{array}{ll}
H^{k,\ell}(\frak B,d) &= 0 \ \ \mbox{if $\ell \neq 0$ or if
$k$ is odd}\\
H^{2n,0}(\frak B,d) & = \{ P(F) \mid P \in \cal
I^n_S(\frak g)\} \simeq \cal I^n_S(\frak g)
\end{array}
\right. \]
and
\[ \left\{ \begin{array}{ll}
H^{k,\ell}(\frak B,\delta) &= 0 \ \ \mbox{if $k \neq 0$
or if $\ell$ is odd}\\
H^{0,2n}(\frak B,\delta) & = \{ P(\varphi) \mid P \in \cal
I^n_S(\frak g)\} \simeq \cal I^n_S(\frak g).
\end{array}
\right. \]
\\

\noindent{\bf Proof.} By the lemma 2.2, $d'_1$,
(resp. $\delta'_1$), gives an homotopy for $d$,
(resp. $\delta$), for terms which contain $(\psi, \varphi,
B,\beta)$ (resp. $(F,\psi, B,\beta)$), so we are left with
$P(F)$, (resp. $P(\varphi))$ with $P\in \cal I^n_S(\frak g)$.
These cocycles are classically cohomologically independent, (in
fact they describe the basic cohomologies of the Weil
algebras $W_d(\frak g)$ and $W_\delta(\frak g))$.
$\square$\\

\noindent {\bf Remark.} One has with obvious
notations, for $P\in \cal I^n_S(\frak g)$
$$d P(\varphi,\cdots, \varphi) =$$
$$= nP(\varphi,\cdots,\varphi,d\varphi +
[A,\varphi])=nP(\varphi, \cdots, \varphi,\beta)= \delta
(-nP(\varphi,\cdots,\varphi,\psi))$$
and
$$\delta
P(F,\cdots,F)=nP(F,\cdots,F,B)=d(-nP(F,\cdots,F,\psi))$$ which
means that  $$d(H(\frak B,\delta))=0 \ \ \mbox{and}\  \delta
(H(\frak B,d))=0$$  or equivalently,  $$H(H(\frak
B,\delta),d)=H(\frak B,\delta) = \{P(\varphi)|P\in \cal
I_S(\frak g)\} \simeq \cal I_S(\frak g)$$ and
$$H(H(\frak B,d),\delta)=H(\frak B,d) =
\{P(F)|P\in \cal I_S(\frak g)\} \simeq \cal
I_S(\frak g).$$\\

\noindent Finally, let us notice that one has:

$$(d+\delta)F+[A+\alpha,F]=B$$
$$(d+\delta)(F+\psi+\varphi)+[A+\alpha,F+\psi+\varphi]=0$$
$$(d+\delta)\varphi + [A+\alpha,\varphi]=\beta$$
$$(d+\delta)B+[A+\alpha,B]=-[F,\psi]+[\varphi,F]$$
$$(d+\delta)\beta+[A+\alpha,\beta]=-[\varphi,\psi]+[F,\varphi].$$

\noindent Therefore one can apply the same method as in the proof
of theorem 2.1, by changing $(F,\psi,\varphi,B,\beta)$ into
$(F,F+\psi+\varphi,\varphi,B,\beta)$, to prove the following
theorem.
\newpage
\begin{theorem}
The $d+\delta$ cohomology of $\frak B$ is given by
\end{theorem}

\[ \left\{ \begin{array}{l}
H^m(\frak B,d+\delta)=0 \ \mbox{if}\  m\  \mbox{is odd}\\
H^{2n} (\frak B,d+\delta)=\{P(F+\psi+\varphi) | P\in \cal
I^n_S(\frak g)\}\simeq \cal I^n_S(\frak g)
\end{array}
\right. \]

\noindent The $d+\delta$ cohomology is of course only graded by
the total degree. Notice that $A+\alpha$ and $F+\psi+\varphi$
generate a third Weil algebra $W_{d+\delta}(\frak g)$ with
differential $d+\delta$.

\section{The $\delta$ cohomology modulo $d$ of $\frak
B(\frak g)$}

The computation of the $d$ and the $\delta$ cohomologies
of $\frak B$ was a necessary step for the computation of
the $\delta$ cohomology modulo $d$ of $\frak B$ which, as
will become clear later on, contains all relevant
cohomological informations on $\frak B$.

\begin{theorem}

One has $H^{k,\ell}(\frak B,\delta\
mod(d))=H^{k,\ell}(\frak B, d\  mod(\delta))=0$ if $k+\ell$
is odd and $H^{k,\ell}(\frak B,\delta\  mod(d)) \simeq
H^{k,\ell}(\frak B,d\  mod(\delta)) \simeq \cal I^n_S(\frak
g)$ if $k+\ell=2n$. More precisely, on $\frak B$, a
complete system of cohomologically independent $\delta$
cocycles modulo $d$ in bidegree $(m,2n-m)$ which is also
a complete system of cohomologically independent $d$
cocycles modulo $\delta$ is given by:
$$\{ term\ of\  bidegree (m,2n-m)\ in\ P(F+\psi+\varphi) |
P\in \cal I^n_S(\frak g)\}.$$
\end{theorem}

\noindent{\bf Proof.} Let $Q^{k,\ell}$ be an element
of $Z^{k,\ell}(\frak B, \delta\  mod(d))$ and let us denote
by $[Q^{k,\ell}]$ its class in $H^{k,\ell}(\frak B,
\delta\ mod(d))$. By definition there is a $Q^{k-1,\ell+1}
\in\frak B^{k-1,\ell+1}$ such that $\delta
Q^{k,\ell}+dQ^{k-1,\ell+1}=0$. By applying $d$ one gets
$\delta dQ^{k,\ell}=-d\delta Q^{k,\ell}=0$ which implies,
since $H^{k+1,\ell}(\frak B,\delta)=0$, that there is a
$Q^{k+1,\ell-1}\in \frak B^{k+1,\ell-1}$ such that
$\delta Q^{k+1,\ell-1}+dQ^{k,\ell}=0$. If
$[Q^{k,\ell}]=0$, i.e. if $Q^{k,\ell}=\delta L^{k,\ell-1}
+ dL^{k-1,\ell}$ with $L^{k,\ell-1}, L^{k-1,\ell}\in
\frak B$, then one has $\delta(Q^{k+1,\ell-1}
-dL^{k,\ell-1})=0$ and therefore, since $H^{k+1,\ell-1}(\frak
B,\delta)=0$, there is a $L^{k+1,\ell-2}\in \frak B$ such
that $Q^{k+1,\ell-1}=\delta
L^{k+1,\ell-2}+dL^{k,\ell-1}$. Thus $[Q^{k,\ell}]=0$
implies $[Q^{k+1,\ell-1}]=0$ and therefore there is a
well defined linear mapping $\partial':H^{k,\ell}(\frak
B,\delta\  mod(d)) \rightarrow H^{k+1,\ell-1}(\frak
B,\delta\  mod(d))$ defined by
$\partial'[Q^{k,\ell}]=[Q^{k+1,\ell-1}]$.

$\partial'$ {\it is injective for} $(k,\ell)\neq(2n,0)$.
Indeed, assume $\partial'[Q^{k,\ell}]=0$ or, equivalently
$Q^{k+1,\ell-1}=\delta L^{k+1,\ell-2}+dL^{k,\ell-1}$ with
$L^{k+1,\ell-2}, L^{k,\ell-1}\in \frak B$; then
$dQ^{k,\ell}=d\delta L^{k,\ell-1} \ (=-\delta d
L^{k,\ell-1})$ which implies $[Q^{k,\ell}]=0$ if
$(k,\ell)\neq (2n,0)$ since then $H^{k,\ell}(\frak B,d)=0$.

$\partial'$ {\it is surjective}. Indeed let $Q^{k+1,\ell-1}$
be a $\delta$ cocycle modulo $d$ of $\frak B$, i.e. there is
a $Q^{k,\ell}\in \frak B^{k,\ell}$ such that
$\delta Q^{k+1,\ell-1}+dQ^{k,\ell}=0$; then
$d\delta Q^{k,\ell}=0$ which implies that $Q^{k,\ell}$ is also
a $\delta$ cocycle modulo $d$ of $\frak B$ since
$H^{k,\ell+1}(\frak B,d)=0$ and therefore one has
$[Q^{k+1,\ell-1}]=\partial'J[Q^{k,\ell}]$.\\
We have proved that $\partial':H^{k,\ell}(\frak B, \delta\
mod(d)) \rightarrow H^{k+1,\ell-1}(\frak B,\delta \ mod(d))$
{\it is an isomorphism for} $(k,\ell)\neq (2n,0)$, which
implies that
\[ \begin{array}{cc}
\cal I^n_S(\frak g)=H^{0,2n}(\frak B, \delta) \buildrel
\partial' \over \simeq &H^{1,2n-1}(\frak B, \delta\  mod(d))
\buildrel \partial' \over \simeq \cdots\hfill\\
&\cdots \buildrel
\partial' \over \simeq H^{2n-1,1}(\frak B, \delta\  mod(d))
\buildrel \partial' \over \simeq H^{2n,0}(\frak B,d)
\end{array} \]
where, for the last isomorphism, we use $H^{2n,0}(\frak
B,\delta\  mod(d)) \simeq H^{2n,0}(\frak B,d)$ which is implied
by the remark of Section 2.\\
One can use a similar argument for $H(\frak B, d\
mod(\delta))$.\\
Finally, if $P\in \cal I^n_S(\frak g)$, then
$(d+\delta)P(F+\psi+\varphi)=0$ follows from \linebreak[4]
$(d+\delta)(F+\psi+\varphi)=[F+\psi+\varphi, A+ \alpha]$ and
therefore if one defines $Q^{k,\ell}=$ term of bidegree
$(k,\ell)$ in the expansion of $P(F+\psi+\varphi)$ for
$k+\ell=2n$, one has $\delta Q^{k,\ell} + dQ^{k-1,\ell+1}=0$,
i.e. $[Q^{k,\ell}]=\partial'[Q^{k-1,\ell+1}]$.$\square$
\newpage

\begin{center}
\large Part II\\
{\bf CHARACTERISTIC HOMOMORPHISM}\\
Generalisation of the Weil homomorphism
\end{center}

\vspace{1cm}

\setcounter{section}{0}
\section{General framework}

Let $M$ be a smooth finite dimensional manifold and
$P\rightarrow M$ be a smooth principal bundle over $M$ with
structure group $G$ such that $Lie(G)=\frak g$. The gauge
group (vertical automorphisms) $\cal G$ of $P$ acts on the
affine space $\cal C$ of connections on $P$. Let $\cal P
(\subset \cal C)$ be a $\cal G$-invariant smooth manifold of
connections on $P$. On the data $(\cal P, \cal G$) we assume
the following regularity condition: The quotient $\cal M=\cal
P/\cal G$, i.e. the set of orbits, is a smooth manifold in
such a way that $\cal P\rightarrow \cal M$ is a smooth
principal $\cal G$-bundle.\\
For instance one can take $\cal P$ to be the space of
irreductible connections on $P$ and $\cal G$ to be the group
of all vertical automorphisms of $P$, or one can take $\cal
P$ to be the space of all connections on $P$ and $\cal G$ to
be the group of pointed gauge transformations, i.e. the group
of vertical automorphisms leaving invariant one point, and
therefore one fibre, of $P$, (all this with appropriate
smooth structure). Another classical example is, for $M$= a
compact connected oriented 4-dimensional riemannian manifold
and $G$ compact (e.g. $G=SU(2)$), to take $\cal P$ to be the space of
irreductible self-dual connections; in this case $\cal M$ is
a finite dimensional manifold. In any case $\cal M$ plays the
role of a moduli space of connections on $P$.\\
The algebra $\Omega(P\times\cal P)$ of differential forms on
$P\times \cal P$ is a bigraded differential algebra with
differential $d+\delta$ where $d$ is the exterior
differential of $P$ and $\delta$ is the exterior differential
of $\cal P$; $d$ is of bidegree (1,0) and $\delta$ is of
bidegree (0,1). Tangent vectors to $P\times \cal P$ split in
two parts, $X=X^{1,0}+X^{0,1}$, where $X^{1,0}$ is tangent to
$P$ and $X^{0,1}$ is tangent to $\cal P$.\\
There is {\it a canonical Lie $(G)$-valued one form}
$\underline{A}$ on $P\times \cal P$ of bidegree (1,0), i.e.
$\underline{A}\in \frak g \otimes \Omega^{1,0}(P\times \cal
P)$, defined by $\forall (\xi,a)\in P\times \cal P$,
$\underline{A}(\xi, a)$ is the connection form at $\xi\in P$
of the connection $a\in \cal P$, ($a$ is a connection on $P$).\\
The algebra $\Omega(M\times \cal M)$ of differential forms on
$M\times \cal M$ is canonically a bigraded differential
subalgebra of $\Omega(P\times \cal P)$ which we shall
identify as the base of two operations on $\Omega(P\times
\cal P)$.\\
The structure group $G$ of $P$ acts unambiguously on $P\times
\cal P$ via the right action on $P$. There is a corresponding operation, in the
sense of
H. Cartan, of $\frak g=Lie(G)$ in $\Omega(P\times \cal P)$:
For $X\in \frak g$, $i_X$ denotes the inner antiderivation of
$\Omega(P\times \cal P)$ by the vertical vector field of type
(1,0) corresponding to $X$ (infinitesimal action of $G$). One
has $i_X\underline{A}=X$ and $i_X\delta + \delta i_X=0$ so
the corresponding Lie derivation $L_X$ is given by $L_X=i_Xd
+ di_X$ and $L_X\underline{A}=[\underline{A},X]$.\\
There are several possible actions of $\cal G$ on $P\times
\cal P$: One is the simultaneous action on $P$ and on $\cal
P$, another one is the action on $\cal P$ only. For the sequel (i.e. the
characterization of $\Omega(M \times {\cal M})$ in $\Omega(P\times {\cal
P})$), one can use the operation corresponding to any of these two
actions. The simultaneous action might seem more natural since, after
all, ${\cal G}$ is defined by its action on $P$ (vertical automorphisms).
Nevertheless  we choose the
second one because the corresponding operation seems to us easier
to describe; in this case, the infinitesimal action of $\cal
G$ corresponds to vertical vector fields of type (0,1). The
Lie algebra $Lie(\cal G)$ identifies with $Lie(G)$-valued
equivariant functions on $P$, $\frak X:P\rightarrow \frak
g=Lie(G)$. For $\frak X \in  Lie(\cal G)$, $\frak i_{\frak
X}$ denotes the inner antiderivation of $\Omega(P\times \cal
P)$ by the vertical vector field of type (0,1) corresponding to
$\frak X$. One has $\frak i_{\frak X}d+d\frak i_{\frak X}=0$ so
the corresponding Lie derivative $\frak L_{\frak X}$ is given
by $\frak L_{\frak X}=\frak i_{\frak X} \delta + \delta \frak
i_{\frak X}$. Furthermore one has $\frak i_{\frak
X}(\underline{A})=0$
and $\frak L_{\frak X}(\underline{A})=d\frak X
+[\underline{A},\frak X]$ with obvious notations, (reminding
that $\frak X$ is a $Lie(G)$-valued function).\\
$\Omega(M\times\cal M)$ is the set of basic elements of
$\Omega(P\times \cal P)$ for these operations i.e.\linebreak[4] $\cap \{
\ker (i_X) \cap \ker(L_{X'}) \cap \ker(\frak i_{\frak X}) \cap
\ker(\frak L_{\frak X'}) \}$ $X,X'\in Lie(G),\  \frak X, \frak
X' \in Lie\ (\cal G)$.

\section{Characteristic homomorphism}

It is well known that there exist connections on principal
$\cal G$-bundles\linebreak[4] $\cal P\rightarrow \cal M$ for
$G$ compact, (for instance there are such connections
associated to choices of riemannian structures on $M$). In any
case let $\cal A$ be a connection on $\cal P$. The
corresponding connection form is a $Lie(\cal G)$-valued
one-form on $\cal P,$. Remembering that $Lie(\cal G)$ consists
of $Lie(G)$-valued functions on $P$, it follows, by evaluation
on $P$, that this connection form defines a $Lie(G)$-valued
one-form $\underline{\alpha}$ on $P\times\cal P$ of bidegree
(0,1), i.e. $\underline{\alpha} \in \frak g \otimes
\Omega^{0,1}(P\times\cal P)$. One has $\frak i_{\frak
X}(\underline{\alpha})=\frak X$ and $\frak L_{\frak
X}(\underline{\alpha})= [\underline{\alpha},\frak X]$ for
$\frak X \in Lie(\cal G)$ and $i_X(\underline{\alpha})=0$ and
$L_X(\underline{\alpha})=[\underline{\alpha},X]$ for $X \in
Lie(G)$.\\
By the lemma 1.2 of part I, there is a unique homomorphism of
bigraded differential algebras $h:\frak A \rightarrow
\Omega(P\times \cal P)$ such that $h(A)=\underline{A}$ and
$h(\alpha)=\underline{\alpha}$. The following lemma is the
justification of the name base of $\frak A$ for the
subalgebra $\frak B$.

\begin{lemma}

One has: $h(\frak A)\cap \Omega (M\times \cal M) = h(\frak
B)$.

\end{lemma}

\noindent {\bf Proof.} One has
$i_X(h(A))=i_X(\underline{A})=X$ for $X\in Lie(G)$ and $\frak
i_{\frak X}h(\alpha)=\frak i_{\frak
X}(\underline{\alpha})=\frak X$, $\frak i_{\frak X}
h(\xi)=\frak i_{\frak X}(\delta \underline{A})=\frak L_{\frak
X}(\underline{A})=d\frak X + [\underline{A},\frak X]$,
\mbox{for} $\frak X \in Lie(\cal G)$. Therefore, by
horizontality, $h(\frak A)\cap \Omega (M\times \cal M)$ cannot
contain expressions depending on $h(A)$, $h(\alpha)$ and
$h(\xi)$; thus elements of $h(\frak A)\cap \Omega (M\times \cal
M)$ are of the form $h(\cal I(F,\psi,\varphi,B,\beta))$. On
the other hand , if $Z=h(F),h(\psi),h(\varphi),h(B)$ or
$h(\beta)$ one has $L_XZ=[Z,X]$ for $X\in Lie(G)$ and $\frak
L_{\frak X}Z=[Z,\frak X]$ for $\frak X\in Lie(\cal G)$,
therefore invariance of $h(\cal I(F,\psi,\varphi,B,\beta))$ is
equivalent to total ad($\frak g$)-invariance of $\cal I$, i.e.
$h(\frak A)\cap \Omega(M\times \cal M)=h(\frak B)$.$\square$\\
It follows from this lemma that $h$ induces homomorphisms of
the various cohomologies of $\frak B$ in the corresponding
cohomologies of $\Omega(M\times \cal M)$, in particular, $h$
induces an homomorphism $h^c$ of $H(\frak B,\delta\  mod(d))$
in the $\delta$ cohomology modulo $d$ of $\Omega(M\times \cal
M)$. Of course $h$ depends on $\cal A$, however one has the
following result.

\begin{theorem}
The homomorphism $h^c$ is independent of $\cal A$.
\end{theorem}

\noindent {\bf Proof.}
Let $\cal A_t$, $t\in [0,1]$, be a smooth family of
connections on $\cal P\rightarrow \cal M$ and let $h_t$ be
the corresponding family of homomorphisms of bigraded
differential algebras of $\frak A$ in $\Omega(P\times \cal
P)$. One has for $P\in \cal I^n_S(\frak g)$:
$$
\frac{d}{dt} h_t(P(F+\psi+\varphi,\cdots,F+\psi+\varphi))=\hfill$$
$$(d+\delta)(nP(\frac{d}{dt} h_t(\alpha),h_t(F+\psi+\varphi),
\cdots,h_t(F+\psi+\varphi))).$$

\noindent On the other hand
$nP(\frac{d}{dt}h_t(\alpha),h_t(F+\psi+\varphi),
\cdots,h_t(F+\psi + \varphi))$ is basic for the operations of
$Lie(G)=\frak g$ and of $Lie(\cal G)$ and therefore it is an
element of $\Omega(M\times \cal M)$ so, by integration one has
$$h_1(P(F+\psi+\varphi)) - h_0(P(F+\psi+\varphi)) \in
(d+\delta)\Omega (M\times \cal M)$$ which means, in view of
the theorem 3.1 of part I that $h^c_1 - h^c_0=0$.$\square$\\
The homomorphism $h^c$ will be called {\it the characteristic
homomorphism}.\\

\noindent {\bf Remarks.}
\begin{description}
\item{a)}
The same proof shows that the homomorphism $\tilde h^c$ of
$H(\frak B, d\ mod(\delta))$ in the $d$ cohomology modulo
$\delta$ of $\Omega(M\times \cal M)$ is independent of the
choice of $\cal A$. This also shows that the homomorphism
$h^c_{d+\delta}$ of the $d+\delta$ cohomology of $\frak B$ in
the total de Rham cohomology of $M\times \cal M$ does not
depend on $\cal A$. Of course $h^c$, $\tilde h^c$ and
$h^c_{d+\delta}$ are essentially the same thing in view of
theorems 2.2 and 3.1 of part I.

\item{b)}
Similarily, $h^c_\delta:H(\frak B,\delta)\rightarrow H(\Omega
(M\times \cal M),\delta)$ is independent of $\cal A$ because
$$\frac{d}{dt} h_t (P(\varphi))=\delta (nP (\frac {d}{dt} h_t
(\alpha),h_t(\varphi),\cdots,h_t(\varphi)))$$
 and
$$nP(\frac{d}{dt} h_t(\alpha), h_t(\varphi),\cdots,h_t(\varphi))
\in \Omega (M\times \cal M).$$

\item{c)} Finally, $h^c_d:H(\frak B,d) \rightarrow
H(\Omega(M\times \cal M),d)$ is independent of $\cal A$
since $h(P(F))$ is already independent of $\cal A$. In fact
$h^c_d$ is the Weil homomorphism for $P\rightarrow M$:
$$h^c_d:H(\frak B,d)\simeq \cal I_S(\frak g) \buildrel ch
\over \rightarrow H(M)=H(\Omega(M),d) \subset
H(\Omega(M\times \cal M),d),$$
where $ch$ is the usual Chern character of $P\rightarrow M$.

\item{d)}
The Weil homomorphism $ch\simeq h^c_d$ is already included in
$h^c$ since\linebreak[4] $h^c \restriction H^{\ast,0}(\frak
B,\delta\ mod(d)) \simeq h^c_d$. In the same
way,$$h^c\restriction H^{0,\ast}(\frak B, \delta\ mod(d)) \simeq
h^c_\delta$$ so the characteristic homomorphims $h^c$ contains
all the relevant informations and generalises the Weil
homomorphism in an obvious sense.

\end{description}

\section{Cartan operations in $\frak A(\frak g)$ and
$h(\frak A(\frak g))$}

Let us denote by $\underline{Z}$ the image by $h$ of the
generator $Z=A,F,\alpha,\varphi,\psi,\xi,B,\beta$ of $\frak
A$, i.e. $\underline{Z}=h(Z)$. Then the Cartan operations of
$\frak g=Lie(G)$ and of $Lie(\cal G)$ satisfy the following
relations: $(X\in \frak g, \frak X \in Lie(\cal G))$
$i_X(\underline{A})=X$ and $i_X(\underline{Z})=0$ for the other $Z$,
$(\delta X=dX=0)$, $i_X\delta + \delta i_X=0$ so $L_X=i_Xd+di_X$ and
$L_X(\underline{Z})=[\underline{Z},X]$ for
$Z=A,F,\alpha,\cdots,\beta$; \\
$\frak i_{\frak X}(\underline{\alpha})=\frak X$, $\frak i_{\frak X}
(\underline{\xi})=d\frak X + [\underline{A},\frak X]$ and $\frak
i_{\frak X}(\underline{Z})=0$ for the other $Z$,\linebreak[4]
$\frak i_{\frak X}d + d \frak i_{\frak X} = 0$, ($\delta \frak X=0$),
so $\frak L_{\frak X}=\frak i_{\frak X}J\delta + \delta \frak i_{\frak
X}$ and then $\frak L_{\frak X}(\underline{A})=d{\frak
X}+[\underline{A},\frak X]$ and $\frak L_{\frak
X}(\underline{Z})=[\underline{Z},\frak X]$ for
$Z=F,\alpha,\cdots,\beta$.\\
One thus see that the operation of ${\frak g}=Lie(G)$ corresponds under
$h$ to the operation $X\mapsto i_X$ of ${\frak g}$ in ${\frak A}$ defined
in part I. In contrast, there is a problem to define an analogue in
${\frak A}$ of the operation of $Lie({\cal G})$.The reason is that
$d\frak X$ does not mean anything in $\frak A$. So, in order to
define the analogue of this operation in $\frak A$ one has to add
generators, for instance one may combine $\frak A(\frak g)$ with
the Weil-BRS algebra $\cal A(\frak g)$~\cite{6} but the result is
complicated and the interest of the universal model $(\frak
A,\frak B)$ is its simplicity. It is why we refrain to follow
this way and defined directly the base $\frak B$ of $\frak A$ in
part I.\\
However, in contrast to the operation $X\mapsto i_X$, the
operation $X\mapsto i'_X$ does not correspond to an operation
in $\Omega(P\times \cal P)$ under the homomorphism $h$. Thus
the property of $\frak B$ to be basic for $i'$ (and not only for
$i$) looks accidental in the context used here but it is an indication
that there may be another context in which $(\frak A,\frak B)$ plays the
role of a universal model.\\

\section*{Acknowledgements}

Many ideas of this paper come from informal discussions with R. Stora in
Annecy in October 1992. I thank him very much for this and for
constant encouragement/criticism. I thank J.C. Wallet for explanations
of, still unpublished, nice results of R. Stora, F. Thuillier and
himself and for suggestion of several references. I also thank J.C.
Wallet and T. Masson for carefully reading the manuscript. Finally I
thank L. Baulieu, M. Talon and C.M. Viallet for their kind interest.

\end{document}